\documentclass[aps,prl,twocolumn,superscriptaddress,showpacs,floatfix]{revtex4-1}
\usepackage{graphicx,amsmath}
\pdfoutput=1 

\newcommand{\esqoh}{\mbox{$e^2/h$}}

\newcommand{\Startsubfig}[2]{Figure~\ref{fig:#1}(#2)}
\newcommand{\subfig}[2]{Fig.~\ref{fig:#1}(#2)}
\newcommand{\allfig}[1]{Fig.~\ref{fig:#1}}
\newcommand{\Startallfig}[1]{Figure~\ref{fig:#1}}
\newcommand{\refeq}[1]{Eqn.~\ref{eq:#1}}

\begin{document}


\title{Pseudospin-Resolved Transport Spectroscopy of the Kondo Effect in a Double Quantum Dot}

\author{S. Amasha}
	\email{samasha@stanford.edu}
	\affiliation{Department of Physics, Stanford University, Stanford, California 94305, USA}

\author{A. J. Keller}
	\affiliation{Department of Physics, Stanford University, Stanford, California 94305, USA}

\author{I. G. Rau}
	\altaffiliation{Present address: IBM Almaden Research Center, San Jose, CA 95120}
	\affiliation{Department of Applied Physics, Stanford University, Stanford, California 94305, USA}

\author{A. Carmi} 
	\affiliation{Department of Condensed Matter Physics, Weizmann Institute of Science, Rehovot 96100, Israel}
	
\author{J. A. Katine}
	\affiliation{HGST, San Jose, CA 95135, USA }

\author{H. Shtrikman} 
	\affiliation{Department of Condensed Matter Physics, Weizmann Institute of Science, Rehovot 96100, Israel}

\author{Y. Oreg} 
	\affiliation{Department of Condensed Matter Physics, Weizmann Institute of Science, Rehovot 96100, Israel}
	
\author{D. Goldhaber-Gordon} 
	\altaffiliation{Present address:  Stanford University, Stanford, California 94305}
	\affiliation{Department of Physics, Stanford University, Stanford, California 94305, USA}	
	\affiliation{Department of Condensed Matter Physics, Weizmann Institute of Science, Rehovot 96100, Israel}

\begin{abstract}

  We report measurements of the Kondo effect in a double quantum dot (DQD), where the orbital states act as pseudospin states whose degeneracy contributes to Kondo screening. Standard transport spectroscopy as a function of the bias voltage on both dots shows a zero-bias peak in conductance, analogous to that observed for spin Kondo in single dots. Breaking the orbital degeneracy splits the Kondo resonance in the tunneling density of states above and below the Fermi energy of the leads, with the resonances having different pseudospin character. Using pseudospin-resolved spectroscopy, we demonstrate the pseudospin character by observing a Kondo peak at only one sign of the bias voltage. We show that even when the pseudospin states have very different tunnel rates to the leads, a Kondo temperature can be consistently defined for the DQD system.
  
\end{abstract}


\maketitle


   The Kondo effect is one of the paradigms of correlated electron physics \cite{Hewson1993:KondoToHF}. It describes how itinerant electrons with a degenerate degree of freedom screen a localized state with the same degeneracy. Typically, the relevant degeneracy is spin:  a localized electron is transitioned between degenerate spin states by spin-flip scattering with conduction electrons.  Correlations are established between the localized and conduction electrons, with a many-body spin singlet resulting at low temperatures. This Kondo screening causes a resonance in the local density of states (LDOS) at the Fermi energy, which manifests itself in nanostructures as a zero-bias peak in the conductance \cite{Grobis2006:ReviewKondo}. While Kondo physics is usually associated with spin, nanostructures  allow the realization of the Kondo effect based on orbital degeneracy \cite{Wilhelm2002:SpinlessKondo, Holleitner2004:PseudospinKondo, PJH2005:CNTOrbitalKondo, Schroer2006:1eDDKondo, Makarovski2007:CNTSU4}. The advantage to using an orbital degeneracy is its potential to realize a fully-tunable state-resolved probe of Kondo physics that does not perturb the Kondo correlations, which is not possible in spin-based Kondo systems.
   
   Spin-resolved transport measurements in nanostructures have been achieved using ferromagnetic contacts, leading to spin-dependent tunnel rates \cite{Pasupathy2004:FeKondo, Hauptmann2008:FeContactSpinRev, Calvo2009:KondoFeContacts}.  Unfortunately, these spin-dependent rates also affect Kondo physics \cite{Martinek2003:FeLeadKondo, Martinek2003:FeLeadKondoNRG, Sindel2007:FeLeadsNRG}; moreover, the rates are fixed by the contact design and cannot be tuned. Another approach has been to use a quantum point contact (QPC) as a spin polarizer \cite{Potok2003:PolCurrent} to build up a non-equilibrium distribution, with a spin-dependent Fermi energy \cite{Kobayashi2010:SpinAccumulation}. However, this technique requires a magnetic field that breaks the spin degeneracy necessary for the Kondo effect. 

   We instead realize a tunable state-resolved probe of the Kondo effect using an orbital degeneracy of a double quantum dot (DQD) \cite{Hubel2008:DDKondo, Okazaki2011:SOKondo}, which occurs when the energy for an electron to be in dot 1 is the same as that for being in dot 2. These orbital states can be coherently manipulated as a two-level `pseudospin' system \cite{Fujisawa2004:ChargeQubit, Petersson2010:ChargeQubit}. The advantage of studying a Kondo effect based on pseudospin degeneracy is that by controlling and measuring each of the dots individually, we can characterize the conductance of each pseudospin component \cite{Borda2003:DQDSU4Kondo, Feinberg2004:KondoDD, Carmi2011:SUNinBfield, Busser2012:DDPolCur}.

      In this Letter, we report pseudospin-resolved transport spectroscopy of the Kondo effect based on an orbital degeneracy in a DQD. We first demonstrate spectroscopy of the DQD analogous to standard transport spectroscopy in a single dot, and we use this to observe the zero-bias peak that is the hallmark of Kondo physics. In standard spectroscopy of spin Kondo, a magnetic field splits the Kondo peak so that the conductance at zero-bias is suppressed and the Kondo peaks occur at positive and negative bias. In contrast, pseudospin-resolved spectroscopy of the orbital Kondo effect in a pseudo-magnetic field shows a peak at only one sign of the bias, corresponding to the pseudospin state we are observing. Finally, we demonstrate a single, consistent Kondo temperature can be defined for the entire DQD system.

\begin{figure}
\begin{center}
\includegraphics[width=8.0cm, keepaspectratio=true]{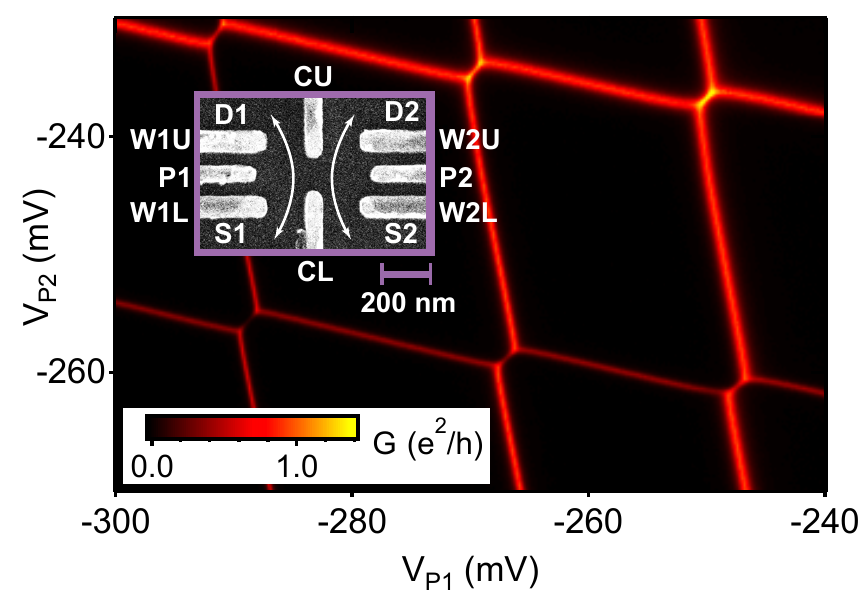}
\end{center}

\caption{(color online) The sum of measured zero-bias conductances through dots 1 and 2 ($G= G_{1}+G_{2}$) as a function of the voltages on the gates labeled P1 and P2 ($V_{\rm P1}$ and $V_{\rm P2}$ respectively).  Inset: A scanning electron micrograph of a device similar to the one measured. The gates that define the DQD as well as the source and drain leads for dot 1 (S1 and D1) and for dot 2 (S2 and D2) are labeled. Arrows have been drawn to emphasize the paths through which currents are measured.
}
\label{fig:largeP1P2}
\end{figure}

      We measure a laterally-gated DQD fabricated from an epitaxially grown AlGaAs/GaAs heterostructure hosting a two-dimensional electron gas (2DEG) with a density of $2 \times 10^{11}~\mbox{cm}^{-2}$ and a mobility of $2 \times 10^6~\mbox{cm}^{2}/\mbox{Vs}$. We apply negative voltages to metallic surface gates (inset to \allfig{largeP1P2}) to form two capacitively-coupled quantum dots with negligible inter-dot tunneling \cite{footnote:EPAPS}. The gates W1L and W1U control the tunneling rates between dot 1 and its source and drain leads $\Gamma_{\rm S1}/ \hbar$ and $\Gamma_{\rm D1}/ \hbar$, respectively. We define $\Gamma_{1}= \Gamma_{\rm S1} + \Gamma_{\rm D1}$, and $\Gamma_{2}$ analogously for dot 2. The conductances of the dots are measured using separate circuits. All the data in this paper are taken with $B\leq 80~\mbox{mT}$, so that spin degeneracy is maintained.
      				
   \Startallfig{largeP1P2} shows the results of summing the zero-bias conductance measured through dots 1 and 2 (denoted $G_{1}$ and $G_{2}$, respectively) as a function of the voltages applied to the gates P1 and P2, which control the occupancy of the dots. The Coulomb blockade lines delineate the ``honeycomb'' shape of the DQD charge stability diagram \cite{vanderWiel2002:DQDreview}. From these data we extract intra-dot charging energies of $U_{1}\approx 1.2~\mbox{meV}$ and $U_{2}\approx 1.5~\mbox{meV}$, as well as an inter-dot charging energy $U^{\prime}\approx 100~\mu\mbox{eV}$.  
   
     For the data in \allfig{largeP1P2}, $\Gamma_{1}$ and $\Gamma_{2}$ are between $20$ and $50~\mu\mbox{eV}$. Since $\Gamma_{1} \ll U_{1}$ and $\Gamma_{2} \ll U_{2}$ the spin Kondo temperature is much less than the electron temperature of $22~\mbox{mK}$ and we do not observe Kondo-enhanced conductance due to spin degeneracy in the odd Coulomb valleys. However, in this regime $\Gamma_{1,2}/U^{\prime}\sim 0.2$ to $0.5$ and between each pair of triple points visible in the figure we observe Kondo-enhanced conductance from an orbital degeneracy \cite{Hubel2008:DDKondo}. In contrast to Ref. \cite{Okazaki2011:SOKondo}, we observe enhancements at all orbital degeneracies, regardless of whether the dots contain an even or odd number of electrons. As spin degeneracy has not been broken it should play a role \cite{Keller:SPSKondo}, but many of the salient features can be explained by considering the orbital degeneracy alone.     
     
\begin{figure}[t!]
\begin{center}
\includegraphics[width=8.0cm, keepaspectratio=true]{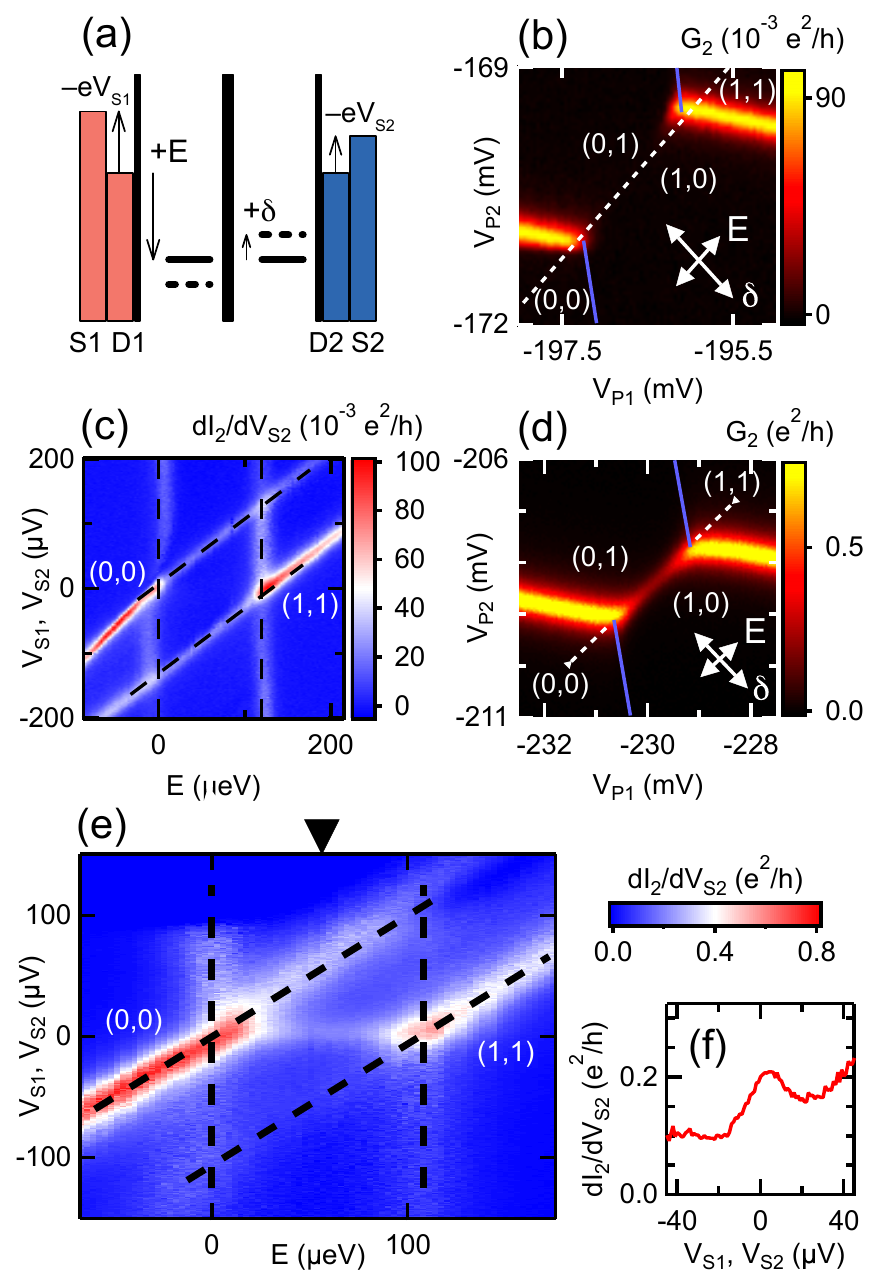}
\end{center}

\caption{(color online)(a) DQD energy diagram showing (from left to right): dot 1's leads, confinement potential, and chemical potential (solid horizontal line). A thick tunnel barrier emphasizes the negligible inter-dot tunneling, and then there is an analogous picture for dot 2. Leads S1 and S2 may be independently biased by voltages $V_{\rm S1}$ and $V_{\rm S2}$. $E$ is the energy of the dot states below the Fermi energy of the drains, while $\delta$ is the energy difference between the dot states (the solid lines show the levels at $\delta=0$ while the dashed lines show how the levels shift with positive $\delta$).  (b) $G_{2}$ for $\Gamma_{1,2}/U\approx 0.13$. Only the dot 2 Coulomb blockade peaks are visible, and the purple lines show where peaks may be observed in transport through dot 1. The ordered pairs list the occupation of the dot states relative to some background occupation. The compass shows the vectors along which $V_{\rm P1}$ and $V_{\rm P2}$ are simultaneously swept in order to change $E$ or $\delta$, and the arrowheads correspond to $\pm25 \mu\mbox{eV}$. The dashed line corresponds to the horizontal axis in \subfig{diamonds}{c} at zero bias. (c) Bias spectroscopy of dot 2 at $\delta= 0$, when the pseudospin states are degenerate. (d) $G_{2}$ in the double dot Kondo regime with $\Gamma_{1,2}/U\approx 0.24$. The compass is as in \subfig{diamonds}{b}, and the dashed line corresponds to the horizontal axis in \subfig{diamonds}{e} at zero bias. (e) Bias spectroscopy for dot 2 at $\delta=0$ in the Kondo regime. The black arrow shows the location of the cut shown in (f).
}
\label{fig:diamonds}
\end{figure}

   To perform the analogue of standard bias spectroscopy on a DQD, we apply an equal voltage to both the pseudospin-up source (S1) and the pseudospin-down source (S2), while varying the energy of the orbital states $E$ and maintaining their degeneracy ($\delta= 0$), see \subfig{diamonds}{a}.  We accomplish this by determining the capacitance factors that relate changes in $V_{\rm P1}$, $V_{\rm P2}$, $V_{\rm S1}$, and $V_{\rm S2}$ to changes in the energies of the dots \cite{footnote:EPAPS}. This allows us to find the gate voltages necessary to effect a change in either the average energy $E$ or the detuning $\delta$ of the double dot system (compass in \subfig{diamonds}{b}) for given bias voltages.

   In the Coulomb blockade regime, standard spectroscopy of the DQD shows the characteristic diamond-shaped regions of suppressed conductance. \Startsubfig{diamonds}{b} shows the conductance through dot 2 when $\Gamma_{1,2}/U^{\prime} \ll1$ and Kondo screening is suppressed, while the corresponding spectroscopy measurements are shown in \subfig{diamonds}{c}. The slopes of the Coulomb diamond edges are as predicted: the vertical dashed lines correspond to alignment of the dot levels with their drain leads while the dashed lines with slope $1$ correspond to alignment with the source leads \cite{footnote:EPAPS}. This agreement demonstrates the high fidelity of our control over $E$ and $\delta$. 
    
      As  $\Gamma_1$ and $\Gamma_2$ are increased, we observe a conductance enhancement along a line between a pair of triple points, where two orbital states are degenerate (\subfig{diamonds}{d}). The corresponding spectroscopy data are shown in \subfig{diamonds}{e} and exhibit a zero-bias peak in the middle of the Coulomb diamond. This provides clear evidence that the conductance enhancement results from Kondo screening. 

\begin{figure}
\begin{center}
\includegraphics[width=8.0cm, keepaspectratio=true]{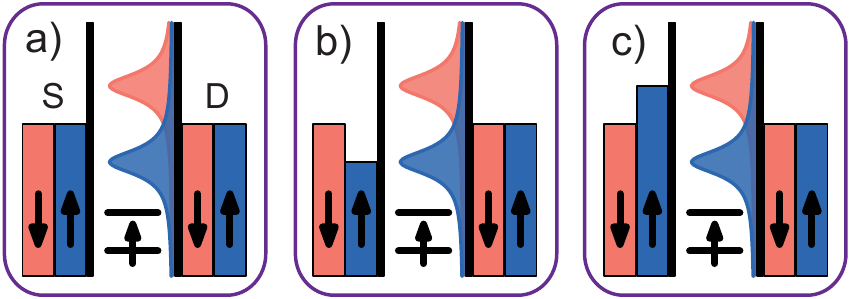}
\end{center}

\caption{(color online) (a) Spin-1/2 Kondo effect for a single dot in a magnetic field. The diagram depicts a spin-resolved source lead, a spin-split Kondo peak in the LDOS (for clarity, we show the peaks associated only with the drain lead), and a spin-resolved drain lead. At zero-bias, no Kondo enhancement is observed. (b) When the spin-up source lead is biased by $V_{\rm S,\uparrow} = +E_{Z}/e$, Kondo-enhanced conductance is observed. (c) For $V_{\rm S,\uparrow} = -E_{Z}/e$, no Kondo enhancement occurs. 
}
\label{fig:KondoInField}
\end{figure}

  To demonstrate pseudospin-resolved spectroscopy, as well as the importance of orbital degeneracy, we can break this degeneracy. We can gain intuition about the results by considering the spin Kondo effect in a single dot in a magnetic field. Above a threshold field, the peak in the LDOS splits above and below the Fermi energy of the leads by the Zeeman energy  $E_{\rm Z}$ \cite{Meir1993:nonequilAnderson, Costi2000:KondoSplitting}. The lower energy peak is associated with spin-up and the higher energy peak with spin-down (\subfig{KondoInField}{a}). At zero bias,  the peaks are no longer aligned with the Fermi energy and no conductance enhancement is observed. The spin-dependent nature of the peaks can be resolved by independently varying the electrochemical potential of one spin species. For example, if the spin-up electrons are biased so that their electrochemical potential aligns with the spin-up peak ($V_{\rm S,\uparrow} = +E_{Z}/e$) then the conductance enhancement should be observed (\subfig{KondoInField}{b}).  Specifically, a spin-down electron can tunnel on from either lead, temporarily violating energy conservation. The spin-up electron can then tunnel out to the source lead, restoring energy conservation and flipping the spin of the dot. This and higher-order spin-flip processes constitute the non-equilibrium Kondo effect. Similar spin-flip processes occur if the spin-down electrons are biased to align with the spin-down peak. In contrast, when the spin-up electrons are biased to align with the spin-down peak ($V_{\rm S,\uparrow} = -E_{Z}/e$), these spin-flip processes are not possible and no enhancement should be observed (\subfig{KondoInField}{c}). 

\begin{figure}
\begin{center}
\includegraphics[width=8.0cm, keepaspectratio=true]{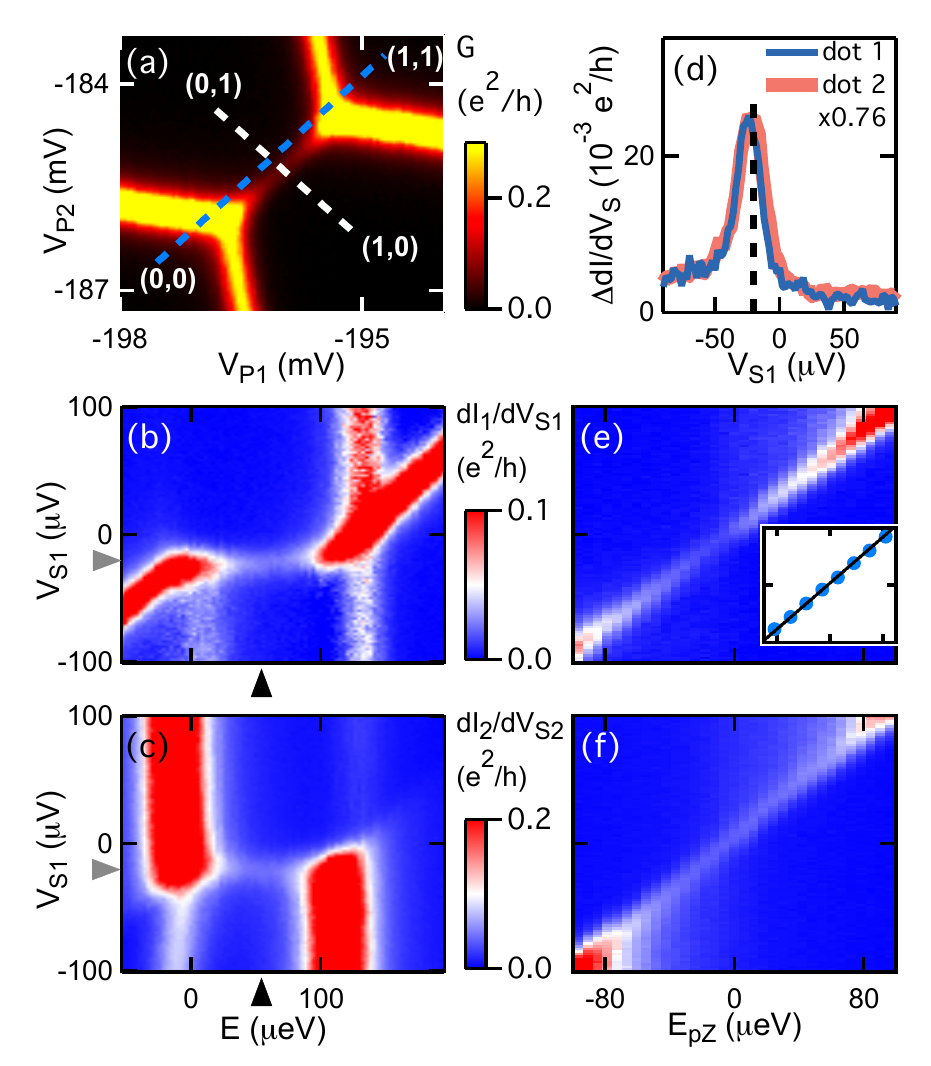}
\end{center}

\caption{(color online) (a)  $G$ as a function of $V_{\rm P1}$ and $V_{\rm P2}$. Along the dashed blue line $E$ changes while $E_{\rm pZ}= -20~\mu\mbox{eV}$. Along the dashed white line $E_{\rm pZ}$ varies while $E$ is constant.  (b) Pseudospin-resolved bias spectroscopy for dot 1. The horizontal axis at zero bias corresponds to the dashed blue line in (a). The gray arrows on the vertical axis marks $V_{\rm S1}= E_{\rm pZ}/e= -20~\mu\mbox{V}$. (c) Pseudospin-resolved bias spectroscopy for dot 2 as a function of $V_{\rm S1}$. (d) Cuts through the data (black arrows in (b) and (c)) explicitly showing the Kondo enhancement. The dashed black line indicates $V_{\rm S1} =-20~\mu\mbox{V}$. The dot 2 data are scaled by 0.76 and a conductance offset is subtracted to allow comparison of the Kondo peak widths. (e) Pseudospin-resolved spectroscopy through dot 1 as $E_{\rm pZ}$ is varied (horizontal axis at zero bias corresponds to dashed white line in (a) ). The inset shows the position of the peak in $V_{S1}$ (vertical axis, limits are $\pm100~\mu\mbox{V}$) as a function of $E_{\rm pZ}$ (horizontal axis, limits are $\pm100~\mu\mbox{eV}$). The solid black line shows the result of fitting to $V_{\rm S1}= E_{\rm pZ}/e+c$ with the offset $c$ as the only fit parameter. (f) Pseudospin-resolved spectroscopy through dot 2 as a function of $V_{\rm S1}$. Since Kondo involves pseudospin flips, the Kondo enhancement observed in dot 1 in (e) gives a corresponding enhancement in dot 2. 
}
\label{fig:PSres}
\end{figure}

   Standard bias spectroscopy of spin Kondo in a single dot in a magnetic field does not resolve the spin-dependent nature of the resonances: the bias changes the electrochemical potential of both spin species so the Kondo enhancement appears at both signs of the bias voltage $V_{S}= \pm E_{Z}/e$ \cite{DGG1998:NaturePaper, Kogan2004:SpinNKondoSplit}. However in a DQD one can perform the pseudospin-resolved measurement by varying the bias on only S1, corresponding to changing the electrochemical potential of the pseudospin-up electrons. To realize this pseudospin-resolved spectroscopy we apply a finite detuning to establish a pseudo-Zeeman splitting $E_{\rm pZ}= 2\delta$ (dashed blue line in \subfig{PSres}{a}, along which $E_{\rm pZ}= -20~\mu\mbox{eV}$). The corresponding pseudospin-resolved bias spectroscopy data are shown for dot 1 in \subfig{PSres}{b}. There is no longer an enhancement at zero bias; rather, we observe the Kondo peak at a finite bias voltage. The peak location is in good agreement with the expected value, $V_{\rm S1}= E_{\rm pZ}/e= -20~\mu\mbox{V}$,  indicated by the gray arrow along the vertical axis. Most importantly, there is no Kondo enhancement at positive bias, demonstrating pseudospin resolution in measurement of the of the Kondo-enhanced density of states. 
      
   As Kondo screening involves pseudospin flips, at $V_{\rm S1}= E_{\rm pZ}/e$ we also expect to see an enhancement in the conductance through dot 2. This is validated in \subfig{PSres}{c}, where we show spectroscopy of dot 2 as a function of $V_{\rm S1}$. \Startsubfig{PSres}{d} shows cuts through the data in \subfig{PSres}{b} and (c) indicated by the black arrows.  As expected, the position of the peaks in $V_{\rm S1}$ agree. We check the dependence of the peak position on $E_{\rm pZ}$, and these data are shown in \subfig{PSres}{e} and (f). The position of the resonance in $V_{\rm S1}$ depends on $E_{\rm pZ}$ as predicted: the extracted positions are shown in the inset to \subfig{PSres}{e}, and the agreement with the solid line demonstrates $V_{\rm S1}= E_{\rm pZ}/e$ up to a small offset. Pseudospin spectroscopy as a function of $V_{\rm S2}$ shows the behavior of the pseudospin-down peak, which has a negative slope as a function of $E_{\rm pZ}$ (see \cite{footnote:EPAPS}).

\begin{figure}
\begin{center}
\includegraphics[width=8.0cm, keepaspectratio=true]{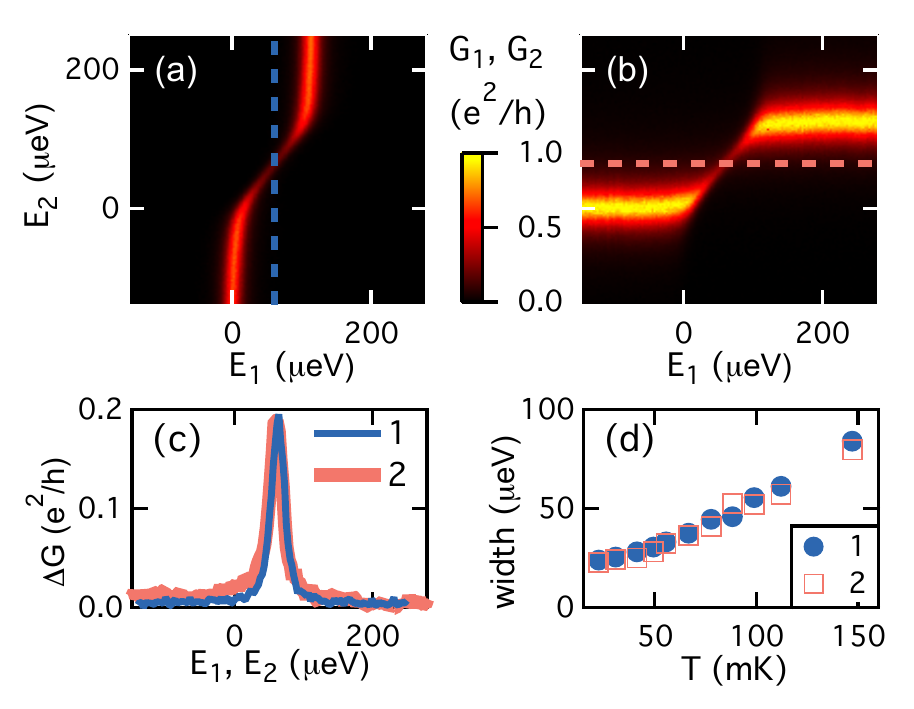}
\end{center}

\caption{(color online) (a) and (b) Zero-bias conductance of dot 1 (a) and dot 2 (b) for $\Gamma_{1}= 19~\mu\mbox{eV}$ and $\Gamma_{2}=45~\mu\mbox{eV}$. The axes refer to the energy of the dot 1 ($E_{1}= E+\delta$) and dot 2 ($E_{2}= E-\delta$) states relative to the drain leads. (c) Conductance cuts through the data indicated by the dashed lines in (a) and (b). To compare the Kondo peak widths a constant conductance background was subtracted from the dot 2 data, but no vertical scaling was necessary. (d) Full width at half-maximum measured from cuts like those in (c) as a function of temperature.
}
\label{fig:AsymRates}
\end{figure}

    The data in \subfig{PSres}{d} show that the widths of the peaks in dot 1 and 2 are equal, indicating that we can define a consistent $T_{K}$ for the DQD. We check that we can continue to define a consistent $T_{K}$ when the pseudospin components have very different couplings to their leads (e.g. $\Gamma_{1} < \Gamma_{2}$),  so that the tunneling rates are pseudospin-dependent. This is analogous to contacting a nanostructure with ferromagnetic leads, although the DQD offers the advantage of probing each pseudospin component independently. \Startallfig{AsymRates} shows data taken when $\Gamma_{2}/\Gamma_{1}\approx 2.4$, and the Kondo enhancement is still observed. Cuts through the data are shown in \subfig{AsymRates}{c} and show good agreement between the peak widths. The temperature dependence of the width shown in \subfig{AsymRates}{d} demonstrates that this agreement is maintained over the entire temperature range measured. These data confirm that a single consistent $T_{K}$ scale can be defined across both pseudospin components, even with very asymmetric coupling. 
    
   In conclusion, we report pseudospin-resolved spectroscopy of a DQD. In a pseudo-magnetic field, we observe a Kondo enhancement at only one sign of the bias, which results from the pseudospin dependence of the split Kondo resonance. We also demonstrate that $T_{K}$ is well defined in the pseudospin system. These measurements demonstrate how for probing the many-body Kondo state DQDs give unique capabilities compared to spin-Kondo systems.

	We are grateful to G. Zarand, C. P. Moca, I.  Weymann, S. E. Ulloa, G. B. Martins, C. A. B\"usser, A. E. Feiguin, and M. R. Calvo for discussions. This work was supported by the NSF under DMR-0906062 and by the U.S.-Israel BSF grant No. 2008149. A.J.K. acknowledges a Stanford Graduate Fellowship and D. G.-G. acknowledges the Weston Visiting Professorship at the Weizmann Institute.

\newpage

\renewcommand{\thefigure}{S\arabic{figure}}
\renewcommand{\theequation}{S\arabic{equation}}
\setcounter{figure}{0}

\noindent \begin{center}{\Large Supplementary Information for Pseudospin Resolved Transport Spectroscopy of the Kondo Effect in a Double Quantum Dot}\par\end{center}{\Large
\par}

\section{Quantitative limits on the inter-dot tunneling}

\begin{figure}[h!]
\begin{center}
\includegraphics[width=8.0cm, keepaspectratio=true]{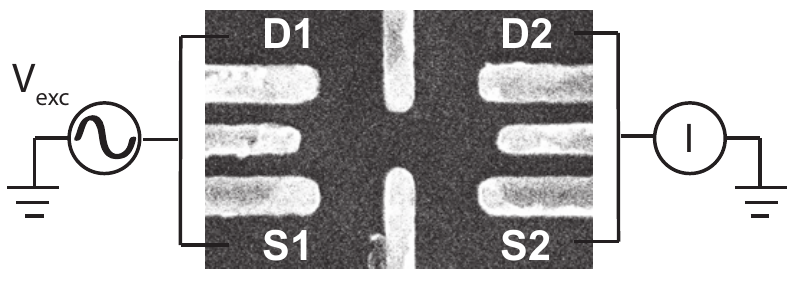}
\end{center}

\caption{Circuit for measuring the series conductance $G_{\rm series}$ of the double quantum dot system as discussed in the text.
}
\label{fig:SeriesMeasurement}
\end{figure}	  

   For all the measurements reported in the main text of this paper, we have made the voltages on the gates labeled CU and CL sufficiently negative so as to suppress inter-dot tunneling. We have confirmed that the inter-dot tunneling is negligibly small over a range of gate voltage settings. To check the inter-dot tunneling, we have measured the conductance of the dots in series, as shown in \allfig{SeriesMeasurement}. When the gate voltages are set to a triple point then three dot states are degenerate, e.g. (0,0), (1,0), and (0,1). At these triple points a finite inter-dot tunneling gives a finite series conductance \cite{vanderWiel2002:DQDreview}. Using the above circuit we have found the conductance at the triple points to be below our measurement threshold, indicating the inter-dot tunneling is small, as detailed quantitatively below.
   
   We can quantitatively limit the inter-dot tunneling energy scale $t$ as follows. At zero-bias, the series conductance $G_{\rm series}$ can be calculated\cite{Nazarov1993:QuantumInterferometer} and is given by:
\begin{equation}
\label{eq:Gseries}
	G_{\rm series}= \frac{64 |t|^2}{3 \Gamma_{1} \left(\frac{\Gamma_{1}+\Gamma_{2}}{2}\right)} \frac{e^2}{h}
\end{equation}
In this equation, $\Gamma_{1}/ \hbar$ is the total tunnel rate between electrons on dot 1 and the leads S1 and D1 (with $\Gamma_{2}$ defined similarly). Using this equation, we can put quantitative limits on $t$.  For gate voltage settings close to those used for the data in Figure 2 of the main text we had $\Gamma_{1}= 44~\mu\mbox{eV}$ and $\Gamma_{2}= 31~\mu\mbox{eV}$ and we found $G_{\rm series} < 4\times 10^{-3} \esqoh$ which gives $|t| < 0.6~\mu\mbox{eV}$. For the data in Figure 4 of the main text, we have  $\Gamma_{1}= 20~\mu\mbox{eV}$ and  $\Gamma_{2}= 22~\mu\mbox{eV}$ and $G_{\rm series} < 3 \times 10^{-3} \esqoh$, which gives a limit $|t|< 0.25~\mu\mbox{eV}$. Both limits on $|t|$ are below the thermal energy scale of $2~\mu\mbox{eV}$ set by our electron temperature of approximately $20~\mbox{mK}$.

\section{Relating changes in dot energies to gate voltages}

  To perform the bias spectroscopy measurements shown in Figures 2 and 4 of the main text, for given values of $V_{\rm S1}$ and $V_{\rm S2}$ we need to adjust the voltages on gates P1 and P2 so as to set the energy $E$ of the orbital states, as well as the energy difference $\delta$ between the states. Achieving this control first requires us to determine the capacitance factors that describe how the voltages $V_{\rm P1}$,$V_{\rm P2}$, $V_{\rm S1}$ and $V_{\rm S2}$ affect the orbital states. We can then use these factors to determine how to vary $V_{\rm P1}$ and $V_{\rm P2}$ so as to control $E$ and $\delta$. This process is described below. 

\begin{figure}
\begin{center}
\includegraphics[width=8.0cm, keepaspectratio=true]{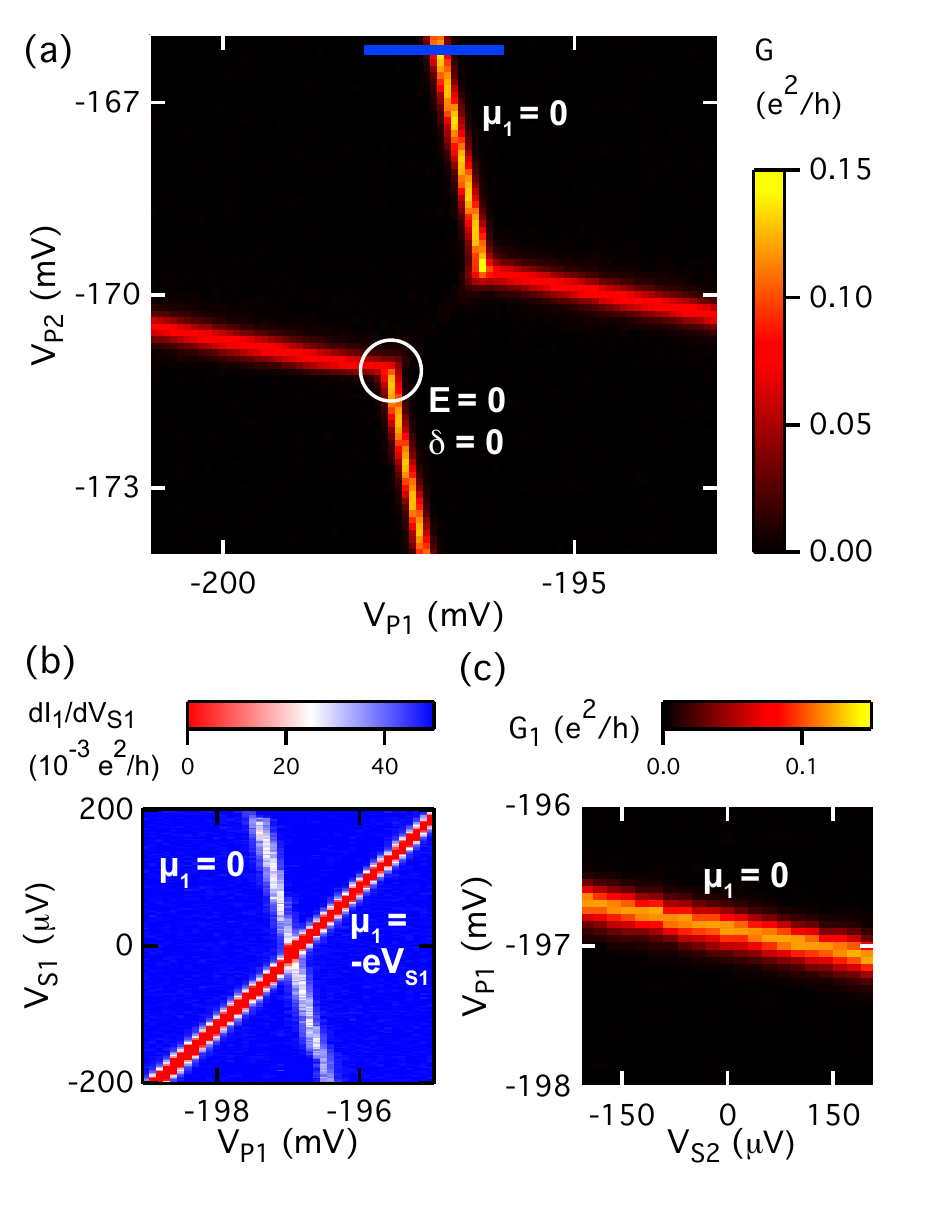}
\end{center}

\caption{(a) Sum of the zero-bias conductance through dots 1 and 2 in the vicinity of a pair of triple points. (b) Bias spectroscopy of dot 1 at $V_{\rm P2}= -166~\mbox{mV}$. The Coulomb diamond associated with the state in dot 1 is clearly visible. The slopes of the diamond edges allow the extraction of capacitance factors as described in the text. (c) Measurement of the zero-bias conductance of dot 1 at $V_{\rm P2}= -166~\mbox{mV}$. The slope of the line allows the extraction of a capacitance factor as described in the text.
}
\label{fig:AlphaMatrix}
\end{figure}

   The electrochemical potential energy of the dot states can be related to the voltages. For dot 1, this relationship can be written as:
\begin{equation}
\label{eq:mu1eq}
	\mu_{1} = -e (\alpha_{\rm P1} \Delta V_{\rm P1} + \xi_{\rm 1,P2} \Delta V_{\rm P2} + \alpha_{\rm S1} V_{\rm S1} + \xi_{\rm 1,S2} V_{\rm S2})
\end{equation}
A corresponding equation holds for dot 2. In this equation, the coefficients of the voltages depend on the capacitance of the gates to the dots\cite{vanderWiel2002:DQDreview} and these factors are what we need to determine. We can directly extract the value of these capacitance factors from the data. \Startsubfig{AlphaMatrix}{a} shows an example of the total conductance for a pair of triple points. To determine the capacitance factors for dot 1, we set $V_{\rm P2}= -166~\mbox{mV}$, where we are away from the charge transition in dot 2 (this position is approximated by the horizontal blue line in \subfig{AlphaMatrix}{a}). At this fixed value of  $V_{\rm P2}$ we perform standard bias spectroscopy of dot 1, as shown in \subfig{AlphaMatrix}{b}. These data show the edges of a Coulomb diamond. Along the edge with negative slope, the dot state is aligned with the Fermi energy of the drain lead, which we assign to be 0. Then we have $\mu_{1}=0$ along this line. Since $\Delta V_{\rm P2}= 0$ and $V_{\rm S2}= 0$, \refeq{mu1eq} gives $\alpha_{\rm P1} \Delta V_{\rm P1} + \alpha_{\rm S1} V_{\rm S1} =0$. So the slope of this diamond edge $m_{i}$ is related to the factors by $m_{i}= V_{\rm S1}/\Delta V_{\rm P1}= -\alpha_{\rm P1}/\alpha_{\rm S1}$. Similarly, the slope of the other diamond edge $m_{j}$ is related to the capacitance factors by $m_{j}= \alpha_{\rm P1}/(1-\alpha_{\rm S1})$. Thus with the measured slopes of the Coulomb diamond, one can extract $\alpha_{\rm S1}$ and $\alpha_{\rm P1}$.

   The remaining capacitance factors can be extracted from other measurements. Along the dot 1 Coulomb blockade line in \Startsubfig{AlphaMatrix}{a} the dot states are aligned with the dot 1 source and drain leads, so $\mu_{1}=0$. Thus from \refeq{mu1eq} we have $\alpha_{\rm P1} \Delta V_{\rm P1} + \xi_{\rm 1,P2} \Delta V_{\rm P2} =0$. This then relates the slope of this line $m_{1}$ to the factors by $m_{1}= \Delta V_{\rm P2} / \Delta V_{\rm P1} = -\alpha_{\rm P1}/\xi_{\rm 1,P2}$. In the data shown in \subfig{AlphaMatrix}{c}, we plot the zero-bias conductance of dot 1 as a function of $V_{\rm S2}$, which gates dot 1. The slope of this line is related to the capacitance factors by $m= -\xi_{\rm 1,S2}/\alpha_{\rm P1}$. In this way, we can extract all the necessary capacitance factors for dot 1. A similar procedure can be used to extract the capacitance factors associated with dot 2. 
   
   The energy of the dot states $E$ and $\delta$ can be related to $\mu_{1}$ and $\mu_{2}$ by 
\begin{displaymath}
	  \begin{pmatrix} E \\ \delta \end{pmatrix} = \frac{1}{2} \begin{pmatrix}\mu_{1} + \mu_{2} \\ \mu_{1}-\mu_{2} \end{pmatrix}
\end{displaymath}
Combining this with \refeq{mu1eq} and a corresponding equation for $\mu_{2}$, we have:
\begin{align}
	 \frac{-2}{e}  \begin{pmatrix} E \\ \delta \end{pmatrix}& = \begin{pmatrix}(\alpha_{\rm P1}+\xi_{\rm 2,P1})&(\alpha_{\rm P2}+\xi_{\rm 1,P2}) \\ (\alpha_{\rm P1}-\xi_{\rm 2,P1})&-(\alpha_{\rm P2}-\xi_{\rm 1,P2}) \end{pmatrix} \begin{pmatrix} \Delta V_{\rm P1} \\ \Delta V_{\rm P2} \end{pmatrix} \notag\\ &+ \begin{pmatrix}(\alpha_{\rm S1}+\xi_{\rm 2,S1})&(\alpha_{\rm S2}+\xi_{\rm 1,S2}) \\ (\alpha_{\rm S1}-\xi_{\rm 2,S1})&-(\alpha_{\rm S2}-\xi_{\rm 1,S2}) \end{pmatrix} \begin{pmatrix} V_{\rm S1} \\ V_{\rm S2} \end{pmatrix}\label{eq:Edelta}
\end{align}
For $V_{\rm S1}= V_{\rm S2}= 0$, we define $E=0$ at the triple point indicated in \subfig{AlphaMatrix}{a}, where the dot states are degenerate ($\delta= 0$) and they are at the same energy as the source and drain leads. For given values of $V_{\rm S1}$ and $V_{\rm S2}$ we can find the necessary gate voltage changes to establish the desired values of $E$ and $\delta$ by substituting into \refeq{Edelta} and solving for $ \Delta V_{\rm P1}$ and $ \Delta V_{\rm P2}$. 
 
 \section{Double Dot Coulomb Diamonds}

\begin{figure}[h!]
\begin{center}
\includegraphics[width=8.0cm, keepaspectratio=true]{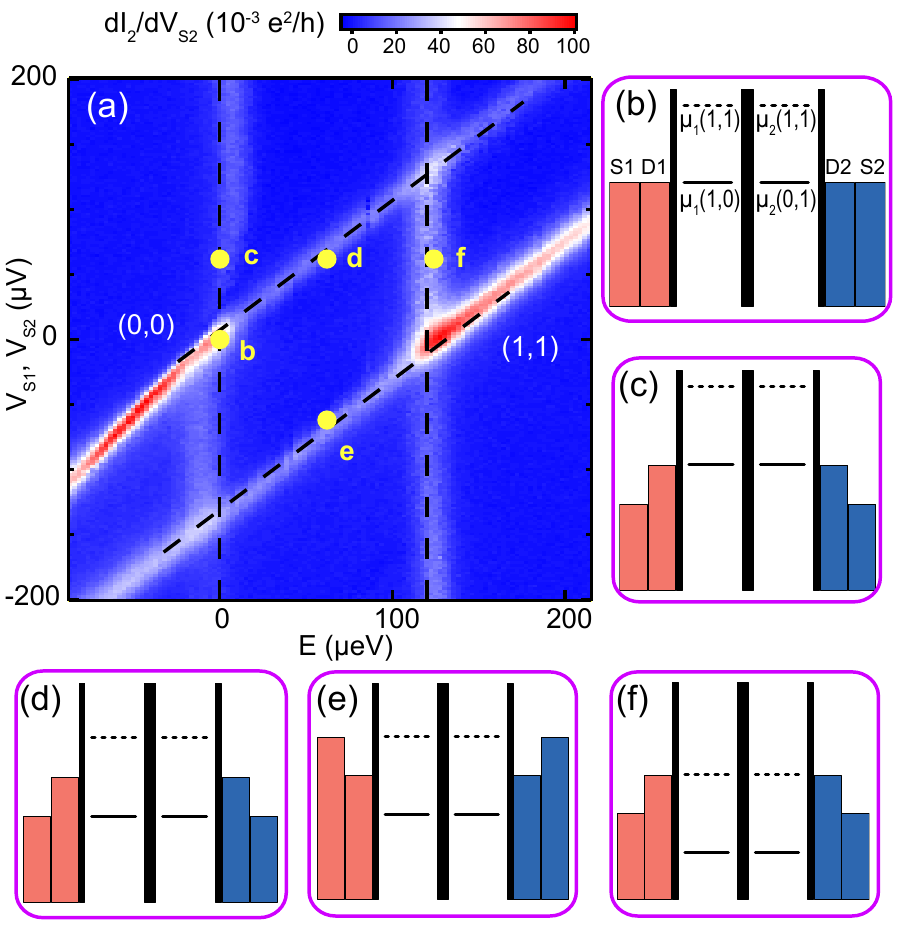}
\end{center}

\caption{(a) DQD Coulomb blockade diamond from Fig 2(c) of the main text. The labeled points correspond to the DQD diagrams in (b) through (f). The dot occupations labeled in (a) and (b) are relative to a background occupation denoted $(0,0)$.
}
\label{fig:DiamondDiagrams}
\end{figure}	  

   Figure 2(c) of the main text shows a Coulomb blockade diamond obtained by performing standard bias spectroscopy at a pair of triple points of the DQD. We can understand the processes that give rise to transport along each edge of the Coulomb diamond.  The data from Fig. 2(c) are shown in \subfig{DiamondDiagrams}{a}, while \subfig{DiamondDiagrams}{b} -(f) show the DQD energy diagrams corresponding to the different points marked in \subfig{DiamondDiagrams}{a}. In these diagrams, the solid lines represent the electrochemical potential energy of the charge states of the DQD (the charge states are labeled relative to some background occupation denoted $(0,0)$); see \subfig{DiamondDiagrams}{b}. For example, $\mu_{1}(1,0)$ denotes the energy for adding an electron to dot 1 when dot 2 contains $0$ electrons. Similarly, $\mu_{2}(0,1)$ is the energy to add an electron to dot 2 when dot 1 contains 0 electrons. The dashed lines  represent the energy to add a second electron to the double dot: for example $\mu_{1}(1,1)$ is the energy to add a second electron to dot 1 when dot 2 contains the first electron.
   
      \Startsubfig{DiamondDiagrams}{b} shows the position of the levels at one of the triple points, when $\mu_{1}(1,0)$ and $\mu_{2}(0,1)$ are degenerate with the Fermi energy of the leads. Along the transport line marked by (c) the positive bias voltages on S1 and S2 lower the electrochemical potential of these leads; however, the dot levels are still aligned with the Fermi energy of the drain leads allowing transport through the dots. Conversely, along the line marked by (d) the dot states  $\mu_{1}(1,0)$ and $\mu_{2}(0,1)$ are below the Fermi energy of the drain leads and transport occurs when they are aligned with the electrochemical potentials of S1 and S2. The diamond edge marked (e) corresponds to applying a negative voltage to S1 and S2 to align the electrochemical potential of these leads with $\mu_{1}(1,1)$ and $\mu_{2}(1,1)$. Finally, the vertical transport line labeled (f) corresponds to  $\mu_{1}(1,1)$ and $\mu_{2}(1,1)$ aligning with the Fermi energy of the drain leads.

\section{Pseudospin Spectroscopy with $V_{\rm S2}$}

   Applying a pseudo-magnetic field splits the Kondo resonance above and below the Fermi energies of the leads by $E_{\rm pZ}$. In Fig. 4 of the main text we resolve the position of the pseudospin-up Kondo peak by sweeping the pseudospin-up lead S1. As expected, the Kondo enhancement of transport through dot 1 follows $V_{\rm S1}= E_{\rm pZ}/e$  (Fig. 4(e) in the main text). Since the Kondo effect involves pseudospin flips, we see a corresponding enhancement of transport through dot 2 when $V_{\rm S1}= E_{\rm pZ}/e$  (Fig. 4(f) in the main text).

\begin{figure}[h!]
\begin{center}
\includegraphics[width=8.0cm, keepaspectratio=true]{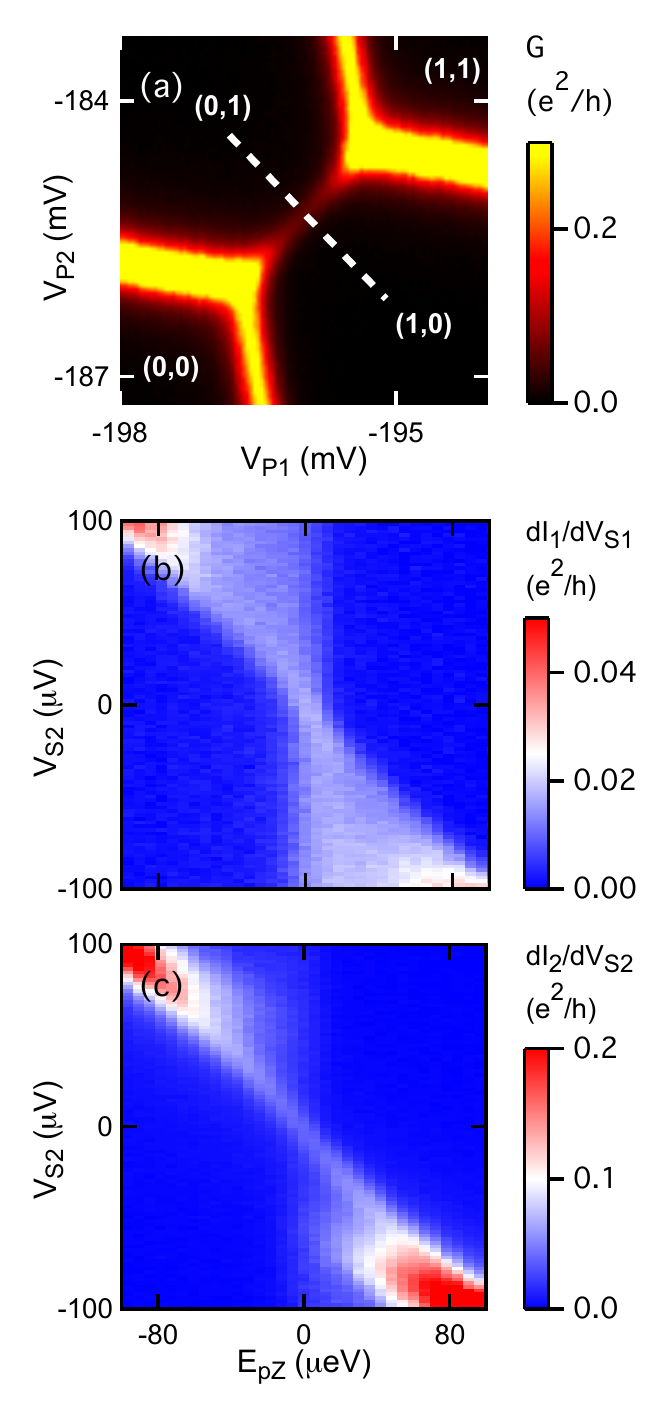}
\end{center}

\caption{(a)   $G$ as a function of $V_{\rm P1}$ and $V_{\rm P2}$. (b) Pseudospin resolved spectroscopy of dot 1 as a function of $V_{\rm S2}$. At $V_{\rm S2}=0$, the horizontal axis corresponds to the dashed white line in (a). (c) Pseudospin resolved spectroscopy of dot 2 as a function of $V_{\rm S2}$. 
}
\label{fig:PSDown}
\end{figure}	  

  To resolve the position of the pseudospin-down peak, which moves in the opposite direction, we need to sweep the pseudospin-down lead S2. \Startsubfig{PSDown}{a}  shows the total conductance $G=G_{1}+G_{2}$ through both dots as a function of $V_{\rm P1}$ and $V_{\rm P2}$. We vary the pseudo-magnetic field by changing the gate voltages to move along the dashed white line in the figure. \Startsubfig{PSDown}{c} shows conductance through dot 2 as a function of the voltage on lead S2 and $E_{\rm pZ}$. Since lead S2 is the pseudospin down lead, this spectroscopy resolves the pseudospin down peak. As expected, we see an enhancement at only one sign of the bias, in this case $V_{\rm S2}= -E_{\rm pZ}/e$. \Startsubfig{PSDown}{b} shows the conductance through dot 1 as a function of $V_{\rm S2}$. As expected, we observe enhanced conductance along the line $V_{\rm S2}= -E_{\rm pZ}/e$. The data in \subfig{PSDown}{b} look different from that in (c) because we also observe enhanced conductance along the vertical line at $E_{\rm pZ}=0$. This is associated with a pseudospin Kondo process with the dot 2 drain lead. Specifically, an electron on dot 1 can tunnel off the dot into D1, and an electron can tunnel onto dot 2 from D2. This maintains energy conservation and results in a pseudospin flip. The electron can then tunnel from dot 2 back to D2, while an electron tunnels back onto dot 1 from S1. This type of process (as well as higher order processes) lead to an enhanced conductance through dot 1, but do not give transport through dot 2. Note that these type of processes rely on strong coupling between dot 2 and D2: the corresponding feature in Fig. 4(f) of the main text is very faint because D1 is only weakly coupled to dot 1.


\newcommand{\noopsort}[1]{} \newcommand{\printfirst}[2]{#1}
  \newcommand{\singleletter}[1]{#1} \newcommand{\switchargs}[2]{#2#1}

\end{document}